\documentclass[10pt,sigconf,nonacm]{acmart}
\renewcommand\footnotetextcopyrightpermission[1]{}
\usepackage{hyperref}
\usepackage{xspace}

\usepackage{ifthen}
\usepackage{soul}
\usepackage{amsmath}
\usepackage[inline]{enumitem}
\usepackage{algorithm}
\usepackage{algpseudocode}

\usepackage{algpseudocode}
\usepackage{etoolbox, xstring}
\usepackage{xspace}
\DeclareListParser{\doslashlist}{/}
\newcounter{ndnNameComponentCounter}%
\newcommand{\name}[1]{{%
		\setcounter{ndnNameComponentCounter}{0}%
		\renewcommand{\do}[1]{{%
				\ifnumgreater{\value{ndnNameComponentCounter}}{0}{\allowbreak/}{}%
				\ifnumodd{\value{ndnNameComponentCounter}}{}{}%
				\detokenize{##1}}%
			\stepcounter{ndnNameComponentCounter}}%
		``{\fontfamily{cmtt}\small\selectfont\IfBeginWith{#1}{/}{/}{}\doslashlist{#1}}''%
}}

\newboolean{showcomments}
\setboolean{showcomments}{true}
\ifthenelse{\boolean{showcomments}}
{ \newcommand{\mynote}[3]{
    \protect\fbox{\bfseries\sffamily\scriptsize#1}
    {\small$\blacktriangleright$\textsf{\emph{\color{#3}{#2}}}$\blacktriangleleft$}}}
{ \newcommand{\mynote}[3]{}}

\usepackage{xcolor}

\definecolor{verylightgray}{gray}{0.8}

\usepackage[normalem]{ulem}

\title{The Internet Runs on Names}

\author{Geoff Huston}
\affiliation{%
\institution{APNIC}
\country{Australia}}
\email{gih@apnic.net}

\author{Lixia Zhang}
\affiliation{%
\institution{UCLA Computer Science Department}
\country{USA}}
\email{lixia@cs.ucla.edu}

\begin{abstract}
The Internet's TCP/IP architecture was designed for resilient packet delivery between hosts identified by IP addresses. Over time, however, the consolidation of applications and services into large-scale platforms built on that universal packet-delivery substrate drove deployment practices that fundamentally changed the Internet's operational model: the network now operates primarily on names. DNS names have become the basis for service identity, reachability, load balancing, and trust, while IP addresses have become ephemeral routing locators. 
This change was driven by application needs and platform consolidation in the absence of any overarching plan.
The resulting mismatch between the original address-based design and the current name-based operation leads to serious consequences: operational complexity that grows with each new layer of indirection, fragility, and vulnerability — as seen in recent high-profile outages.
This paper exposes this mismatch as a necessary first step toward understanding its consequences and addressing the risks of continuing on the same path.
\end{abstract}

\begin{document}
\maketitle

\section{Introduction}
The Internet's original protocol architecture, as described in~\cite{rfc791, RFC793} in 1981, was a principled design. 
Its primary goal was the resilient delivery of packets between hosts across a collection of heterogeneous, independently administered networks. The resulting design choices --- connectionless datagrams, stateless packet forwarding, and the end-to-end principle~\cite{clark88, saltzer84} --- were deliberate: simplicity and resilience at the network layer, with complexity pushed to the edges. IP addresses served as the universal identifiers for communicating hosts; routing was based on destination addresses, steering packets hop-by-hop to their receivers. IP addresses carried the burden of semantic overload, serving both as the unique identifier of an endpoint and as the endpoint's network location. 
The Domain Name System, introduced in 1983~\cite{rfc882, rfc883}, was \emph{not} part of this protocol architecture. It was a lookup convenience layered on top of TCP/IP for humans to find addresses without memorizing numbers. 
The network itself knew nothing of such names.

That clear separation between names and addresses has not endured the pressures of a global, massive, commercial Internet. 
Over the decades, the consolidation of applications and services into large-scale platforms built on that universal packet-delivery substrate has driven deployment practices that have progressively displaced IP addresses as the operational basis of the Internet, replacing them with DNS names.
Today, services are identified, reached, authenticated, and load-balanced by DNS name. 
IP addresses function as ephemeral routing locators assigned for the duration of a connection --- not as stable identifiers for the communicating entities.
Yet the security infrastructure built on top of this name-based 
operation was never grounded in DNS names: each security layer --- 
from routing integrity to application authentication --- was built 
independently, on whatever identifier substrate was available at the 
time, leaving security fragmented across incompatible namespaces 
while operations quietly converged on names.

This transition was not the result of any architectural decision. 
It was the cumulative outcome of independent incremental engineering choices, each made in response to an immediate operational need and the pressures of platform consolidation, service scalability, and service cost minimization, with no overall design. 
The Internet today is neither an address-based architecture designed from the start nor a coherent name-based replacement. 
It is a hybrid: a name-based service layer riding on an address-based forwarding substrate, held together by machinery --- NAT, CDN orchestration, DNS steering, tunneling stacks --- that was never intended at the outset to be load-bearing.

The consequences of this unplanned transition are significant. 
The indirection machinery bridging address-based forwarding and name-based services introduces complexity and fragility that have led to some of the most significant Internet outages in recent years. 

This paper documents the transition and analyzes two of its main consequences, operational complexity and fragility. 
It argues that the Internet has already become name-based in operation, and that this reality must be acknowledged and incorporated into its architecture. 
To that end, this paper takes a necessary first step to expose this mismatch,
understand its risks, and motivate a more principled foundation.

\section{The Original Architecture: Resilient Packet Delivery}
Clark's 1988 paper~\cite{clark88} is explicit about the primary goal of the Internet architecture: effective and resilient communication across a collection of heterogeneous, independently administered networks. The design choices were principled.

\textbf{Datagrams and stateless forwarding.}
The network layer was made as simple as possible: connectionless,
stateless, datagram-based packet delivery. Each packet carried a
destination address; each router made an independent forwarding
decision. No per-flow state, no virtual circuits, no in-network intelligence. 
This simplicity ensured that a network had no state to corrupt; if a router failed, packets were simply rerouted around it.

\textbf{The end-to-end principle.}
Complexity was deliberately pushed to the edges~\cite{saltzer84}. 
The network's role was to deliver packets; reliability, ordering, error
recovery, and other higher-level semantics were the responsibility of
endpoints. This separation kept the network delivery simple and resilient.

\textbf{IP addresses as universal identifiers.}
Every connected host was assigned a globally unique IP address. 
The address was both the network-layer identifier and the routing key: it told the network where to deliver the packet and identified the host to which it was delivered. 

\textbf{DNS as a lookup convenience, not an architectural component.}
The Domain Name System was introduced to spare users from directly working with addresses, mapping human-comprehensible names to IP addresses. 
From the network's perspective, DNS was an invisible application: the network routed addresses; DNS merely helped applications to find these addresses.

This was a clean, coherent design whose weakness lay not in its principles but in its assumptions: that addresses would remain globally unique and stable, and that the network's role would remain pure packet delivery.

\section{The Evolution of Deployment: Name-Based Operation}

The departure from address-based network operation was not planned. 
It was driven, in successive waves, by network growth, application needs, and platform consolidation. Each wave was independently engineered, each adding another layer between the original architecture and operational reality.

\textbf{Address scarcity and NAT.}
The explosive growth of the Internet in the early 1990s led to IPv4 address exhaustion, which led to Network Address Translation (NAT) as a quick, pragmatic response.
NAT lets many devices share either a single IP address or a common pool of IP addresses, with port multiplexing used to demultiplex sessions.
NAT broke global address uniqueness, collapsing the any-to-any host communication model into an asymmetric client/server architecture — most devices became address-anonymous clients; only servers retained stable external addresses. This structural shift pushed applications toward client-initiated connections and server-side identity, accelerating the transition from address-based to name-based operation.

\textbf{Scaling content delivery: CDNs and DNS steering.}
As the Web grew, TCP/IP's point-to-point, address-based packet delivery from a single origin server became a fundamental mismatch: the server's capacity could not meet demand, any origin failure meant total outage, latency was uncontrollable, and global reach required expensive long-haul infrastructure. 
CDNs~\cite{huston25} mitigated this problem by replicating content geographically and using DNS to steer each user to the nearest replica by returning a dynamically selected, location-optimized address rather than a fixed one. 
The DNS name became the stable service identity; the IP address became an ephemeral routing locator.

This inversion --- name as identity, address as ephemeral routing locator --- became the dominant model for large-scale service delivery. 
As we show in Section~\ref{sec:cname}, a DNS lookup for a major web service today typically traverses a chain of CNAME delegations before resolving to an address that is specifically chosen for the user's geography, the current load distribution, or the CDN's real-time health assessment. 
The DNS was not designed for this role; it was repurposed to fill it.

\textbf{HTTP as universal substrate.}
Another shift was the emergence of HTTP and HTTPS as a universal communication substrate, replacing direct IP-based delivery for an increasingly broad range of traffic.
Several forces drove this convergence.
One driver was security: HTTPS provided encryption and server authentication that raw TCP/IP connections lacked.
Another was deployment constraints: firewalls block arbitrary transport protocols and ports, which forced applications to operate over port 443, which is open in virtually every firewall and NAT device.
Additionally, HTTP benefits from a rich operational ecosystem of proxies, load balancers, monitoring tools, and caching infrastructure that no alternative protocol can match, making it the path of least resistance for new applications.
This convergence to HTTP as a universal transport substrate led to the standardization of UDP-over-HTTP~\cite{rfc9298}, IP-over-HTTP~\cite{rfc9484}, and Ethernet-over-HTTP~\cite{ietf-masque-connect-ethernet}, where the entire stack is carried as payload inside HTTPS.
The endpoint's IP address is incidental; the DNS name is the primary session anchor.

\textbf{DNS-based load balancing.}
Beyond CDN steering, DNS has been repurposed into a real-time control plane: load balancing across service instances, geographic distribution of traffic, and health-based failover. These applications treat the DNS as a dynamic, programmable traffic director that returns different answers based on load, system health checks, and policy. 

\textbf{Where we are today.}
The cumulative effect of these shifts is an Internet that operates on names.\begin{verbatim}
Original: Application→IP address→Routing→Host
Today:    Application → DNS name → CDN/DNS
   →dynamic IP→Routing→Service infrastructure
\end{verbatim}
DNS names identify services, anchor TLS sessions, direct CDN traffic, orchestrate load balancing, and serve as the basis for authentication.
The original TCP/IP architecture has not been replaced; it still provides the packet delivery substrate. But the operational model layered on top of it is name-based throughout, and since it is now load-bearing, its failure modes are global in scope.

\section{Consequences: Complexity, Fragility, and Security Fragmentation}
\label{sec:consequences}
The mismatch between the address-based architecture and the name-based operational reality has produced significant, measurable consequences across three dimensions.

\subsection{Operational Complexity}
Bridging address-based forwarding with name-based service delivery requires substantial machinery: NAT traversal, CDN indirection, DNS steering logic, encapsulation stacks for tunneled protocols, and virtualization infrastructure.
This accumulation of compensating mechanisms introduces what we term \textit{indirection debt}: overhead, configuration complexity, and failure points introduced by repeated name–address mappings across layers.
Troubleshooting a simple connectivity issue may require navigating through multiple mapping and tunneling layers.

\subsection{Fragility: DNS Playing Out of Position}
\label{sec:fragility}
The more visible consequence is fragility.
DNS was designed for simple, deterministic lookup: given a name, return an address.
When DNS is used as a real-time control plane, cached data leakage, timing dependencies, race conditions, and cascading failure modes emerge that the original design provided no protection against.

The consequences have been visible at scale. In October 2021, a routine configuration change at Facebook inadvertently caused BGP to withdraw routes to Facebook's authoritative DNS servers~\cite{facebook21}.
Because the entire service ecosystem --- including internal authentication --- relied on DNS for reachability, this single failure made Facebook, Instagram, and WhatsApp globally unreachable for approximately six hours.
The packet-delivery substrate was functional, but the name layer had disappeared.

In October 2025, a software update exposed a race condition in the DNS management system used by Amazon DynamoDB~\cite{aws2025}.
Two independent DNS automation components processed conflicting updates, one overwrote the other, and then deleted the resulting record, leaving the regional DynamoDB endpoint with no DNS resolution.
The failure of this single name entry cascaded to 113 dependent AWS services, causing a 15-hour outage with estimated damages in hundreds of millions of dollars.
Again, IP packet delivery was not the failure; DNS used out of position was.

About one week after the AWS failure, Microsoft Azure experienced an eight-hour global outage triggered by a faulty configuration change in Azure Front Door (AFD), Microsoft's CDN and traffic-routing layer~\cite{azure2025}. A schema validation defect in the deployment pipeline allowed an invalid configuration to propagate globally, disrupting traffic routing for Microsoft 365, Xbox Live, the Azure portal, and thousands of customer applications — all while the underlying packet-delivery substrate remained fully functional.

Three weeks later, on November 18, 2025, Cloudflare's global CDN infrastructure experienced an approximately six-hour outage affecting core content delivery and security services worldwide~\cite{cloudflare2025}. A routine database permissions change caused the automated generation of an oversized bot management configuration file, which was propagated to all edge nodes globally. The traffic routing software on those nodes could not handle the malformed file gracefully and crashed, returning HTTP 5xx errors to users across thousands of dependent services — including X, ChatGPT, Spotify, and Discord — while the underlying packet-delivery substrate remained fully functional.


These four incidents share a common failure mode: a network infrastructure component
repurposed beyond its original design, carrying load-bearing operational
responsibility it was never intended to bear. DNS was designed as a
deterministic lookup service; it has been repurposed as a real-time
traffic director. CDN orchestration was designed to optimize content
delivery; it has become the primary communication channel for a
substantial fraction of Internet traffic. In each case, the
packet-delivery substrate was functional. In each case, the name-based operational layer was the point of failure. And in each case, the blast
radius was bounded not by any architectural property but only by the
reach of each provider's internal dependency graph and their intended service domain.

The failures above were each bounded by their provider's dependency graph, but did not propagate beyond it. 
A second, compounding source of fragility is less visible but potentially broader in scope: the security infrastructure itself has become load-bearing operational infrastructure, with failure modes that propagate not only within a single provider but across provider boundaries. Section~\ref{sec:security} documents this through the case that makes it most concrete.


\subsection{Security: Fragmented and Now Load-Bearing}
\label{sec:security}

Security solutions for the Internet were developed piecemeal, each
driven by immediate demand and built on whatever identifier substrate
was available at the time. The result is a stack of mechanisms that
secure incompatible identifier spaces independently: IP addresses
secured by RPKI~\cite{rfc6480}, DNS mappings secured by
DNSSEC~\cite{rfc4033}, and application sessions secured by Web PKI
and TLS with certificates bound to DNS names. Each layer secured its
own perimeter; no mechanism secures the whole. This fragmentation is
a structural weakness of the name-based operational layer that has
accumulated on top of address-based forwarding --- but it is not the
sharpest consequence. The sharpest consequence is that the security
layer itself has become load-bearing operational infrastructure,
subject to exactly the same failure dynamics as the DNS steering and
CDN orchestration failures documented in
Section~\ref{sec:fragility} --- and with a blast radius that extends
across provider boundaries.

Identity and Access Management (IAM), in a name-based service
architecture, is not a peripheral security control: it is the runtime
gatekeeper through which every service call, every API request, and
every inter-service interaction must pass to obtain an authentication
token. 
On June 12, 2025, Google Cloud's IAM service stopped issuing tokens across more than 40 cloud locations simultaneously~\cite{cloudflare-google2025}, freezing operations rather than merely compromising security.
Every dependent service lost the ability to authenticate,
and the cascade propagated immediately across 26 core Google Cloud
services.

The cascade did not stop at Google's boundary. Cloudflare's Workers
KV service --- critical infrastructure for authentication,
configuration, and asset delivery across Cloudflare's platform ---
depended on Google Cloud storage as its backend. As Google IAM
stopped issuing tokens, Workers KV lost access to its backend and
went offline, taking with it Cloudflare Access, WARP, Workers AI,
and parts of the Cloudflare dashboard for 2.5 hours. Services that
appeared superficially independent shared a hidden dependency and
failed together. This is precisely the dependency that IP-based
delivery does not have: a user reaching a service over IP has no
dependency on the authentication infrastructure of a third-party
cloud provider. A user reaching the same service through a CDN whose
backend depends on that provider does.

This incident closes the argument of Section~\ref{sec:consequences}.
The three failure cases --- DNS steering, CDN orchestration, and
centralized IAM --- are not three separate problems. They are three
instances of the same structural condition: infrastructure repurposed
beyond its original design, now carrying load-bearing operational
responsibility it was never intended to bear, with no isolation
between failure domains. In each case, the packet-delivery substrate
was functional. In each case, the name-based operational layer was
the point of failure. And in each case, the blast radius was bounded
not by any architectural property, but only by the reach of the
hidden dependency graph --- a graph whose full extent, as
Section~\ref{sec:agenda} argues, has not yet been measured.

\section{Toward A Systematic Understanding of Name-based Internet: A Research Agenda}
\label{sec:agenda}

The analysis in Section~\ref{sec:consequences} is grounded in documented
incidents rather than systematic measurement and analysis. 
This reflects a broader gap in the research literature: while the Internet measurement community has produced substantial work on individual components --- DNS resolver behavior and query latency~\cite{moura2020centralization}, CDN mapping and geographic
distribution~\cite{calder2013google}, and third-party service
dependencies~\cite{kashaf2020dependencies} as a few examples --- these studies were conducted in isolation, each focused on a specific layer or mechanism. 
None frames the problem as we do here: a structural mismatch between name-based operation of services and content delivery and address-based packet traversal, and cumulative cost of this mismatch---in indirection complexity, fragility, and hidden dependency---has never been measured as a whole. The overall picture remains unseen, and the cumulative costs unquantified.
Establishing that picture requires a coordinated research agenda across
three areas.

\subsection{DNS as an Encoded Dependency Architecture}
\label{sec:cname}

\noindent\textbf{DNS as a two-level logical routing architecture.}
DNS was originally designed as a deterministic lookup system: given a
name, return the same address to all queriers. This design objective has not
survived contact with CDN-scale deployment. Under pressure from CDN operators
to use the DNS to guide the steering of clients to servers, the DNS has been
progressively repurposed into a two-level logical routing architecture,
in which CNAME chains select \emph{which} operational entity serves a
name, and SVCB records specify \emph{how} to connect to it. 
Neither function was part of the original DNS design; both are now load-bearing infrastructure for name-based service delivery.

\smallskip
\noindent\textit{Level one: CNAME chains as administrative delegation.}
CNAMEs allow individual names to be lifted out of the domain in which
they are syntactically defined and transferred to a different domain
whose operator controls the name's operational behavior. This
mechanism has been repurposed to encode multi-party service
architectures directly in the namespace. The following is a real
example from Akamai's CDN:
{\small\begin{verbatim}
$ dig A www.abc.net.au
www.abc.net.au.     254 IN CNAME
    www.abc.net.au.edgekey.net.
www.abc.net.au.edgekey.net. 2131 IN CNAME
    e3161.b.akamaiedge.net.
e3161.b.akamaiedge.net.   8 IN A  104.83.204.120
\end{verbatim}}

The two CNAME hops perform structurally distinct functions.
The first hop transfers administrative control: it extracts the name
\texttt{www.abc.net.au} from the operational scope of the
\texttt{abc.net.au} domain and hands it to Akamai's \\
\texttt{edgekey.net} subdomain. This record is invariant across all
queriers and carries a high TTL (254 sec), reflecting a stable,
long-term administrative decision by the origin operator.
The second hop performs real-time traffic steering: the response is
specific to the querier's location and carries a TTL of only
8~sec, expiring before a new steering decision may be
needed. 
The final IP address --- selected not by the origin nor by
the user, but by Akamai's real-time orchestration logic --- is an
ephemeral routing locator that may differ for the next query.

The TTL asymmetry is significant beyond caching mechanics. The
intermediate CNAME records define the delegation structure itself ---
which CDN operator controls the name, which edge cluster it routes
to. If these carry short TTLs, the entire routing path can be rewired
in real time, not merely the final address assignment. 
The depth of CNAME chains in operational DNS and the TTL distribution across all hops in those chains --- not only the terminal A~record, whose short TTL is a well-known artifact of CDN steering~\cite{akamai-dns} --- have not been measured systematically. 
Operational deployments routinely exceed two hops --- chaining through provider alias, regional load balancer, and edge cluster --- yet depth and TTL distribution across all hops have not been measured systematically. Doing so, using passive DNS datasets and RIPE Atlas probing, would quantify the extent to which DNS has been repurposed as a control plane.

\smallskip
\noindent\textit{Steering accuracy and ECS.}
The second CNAME hop steers the querier to a topologically proximate
CDN edge node. The authoritative nameserver makes this decision based
on the location of the recursive resolver, not the end user ---
because the DNS query, as standardized, does not carry the
end-user's address. EDNS Client Subnet (ECS)~\cite{rfc7871} was
introduced to address this: it attaches the user's subnet prefix to
the query so that the authoritative nameserver can make a more
accurate steering decision. The accuracy of CDN traffic steering
therefore depends on ECS adoption, which is a deployment choice by
recursive resolvers and CDN operators, not a protocol guarantee.
What has not been measured systematically is how widely ECS is
deployed, how accurately the attached subnet information reflects
actual user location, and how much its absence degrades steering
accuracy for users served by large shared resolvers such as
Google~8.8.8.8 or Cloudflare~1.1.1.1, whose geographic footprint
may be remote from their users.

\noindent\textit{Level two: SVCB records as connection-parameter steering.}
Once CNAME chains have determined \emph{which} infrastructure serves
a name, a second layer of DNS-encoded routing determines \emph{how}
to connect to it. SVCB and HTTPS resource records~\cite{rfc9460}
allow a name's operator to specify, in a single DNS response, the
transport protocol, port, and connection parameters the client should
use --- including ALPN identifiers for HTTP/3 or HTTP/2 preference,
ECH (Encrypted Client Hello) keys, and alternative endpoints. A
simple example: {\small\begin{verbatim}
www.abc.net.au.  HTTPS  1 . alpn=h3,h2 port=443 \
                         ech=AEX... ipv4hint=104.83.204.120
\end{verbatim}}
This record tells the client to prefer HTTP/3 over HTTP/2, use port 443, and attempt an encrypted handshake using the provided ECH key --- all before a single packet has been sent to the server. 
Like the second CNAME hop, the SVCB record is a steering decision encoded in the DNS and controlled by the name's operator. 
Unlike the CNAME chain, which selects the administrative entity responsible for the
name, SVCB selects the logical communication channel through which the connection will be established. 

Together, the two levels form a complete logical routing path: CNAME chooses \emph{who serves the name}; SVCB chooses \emph{how the client connects to them}.
The systematic measurement of SVCB and HTTPS record deployment --- how widely they are published, how consistently clients query for them, and what improvements in connection establishment they produce --- remains an open empirical question.

\subsection{CDN Dominance as Evidence of the Architectural Transition}
CDN infrastructure is not just a performance optimization layered on top of IP delivery; for most users, it is the actual communication channel. Labovitz~\cite{Labovitz2019} measured that approximately 90\% of consumer traffic in 2019 originated from a handful of hypergiants and CDNs. CDNs operate exclusively on DNS names and TLS: a DNS name is the stable service identifier, a TLS certificate bound to that name provides authentication, and IP addresses are assigned transiently by CDN orchestration.

If this traffic share has held or grown --- as industry data suggests~\cite{sandvine2024} --- then CDN traffic dominance is itself the empirical proof of the paper's central claim: name-based communication is the normal steady state of the Internet, not an exceptional condition visible only at failure time. Measuring current CDN traffic share across service categories, using HTTP Archive traces and passive network measurement, would update the 2019 baseline and transform this claim from a qualitative observation into a quantified, category-resolved fact.


\subsection{Resilience of name-based delivery: CDN multihoming.}
In the address-based architecture, resilience against provider failure
is achieved through multihoming: an edge network attaches to two or
more upstream transit providers and announces its address prefixes to
all of them simultaneously. 
BGP provides the standardized machinery for this --- route announcements, traffic engineering attributes, and failover on withdrawal. If a transit provider fails, the prefixes remain reachable through the others.

The name-based delivery layer has no equivalent mechanism.
If a service operator deploys content exclusively through a single CDN, and that CDN fails --- as Cloudflare, Azure Front Door, and Akamai have each done in recent years --- the service becomes unreachable regardless of the health of the IP packet-delivery
substrate. 
An operator wishing to multihome across CDN providers must do so through DNS: publish CNAME records pointing to multiple CDNs and implement health-based failover in the authoritative nameserver. 
But unlike BGP multihoming, there is no standardized protocol or operational framework for this. 
Each operator implements it independently, with no common specification against which correctness or completeness can be verified.

The problem is compounded by hidden backend dependencies that are not visible to content operators. 
A service operator that multihomes across two CDN providers may not know that both depend on a common hyperscaler backend for storage, authentication, or configuration management --- so that a failure in that shared backend propagates simultaneously to both CDNs, and apparent diversity masks actual concentration. Section~\ref{sec:consequences} documents a concrete instance of this dependency structure. 
How widely such hidden convergence exists across the CDN ecosystem, how widely CDN multihoming is practiced despite it, how effectively it provides resilience against both CDN-level and backend-level failures, and where further standards work could provide principled machinery analogous to BGP's role in the address-based layer --- these are open questions whose answers are necessary to understand the resilience of the name-based Internet that has already been built.

Together, these measurements would determine how load-bearing the name layer has actually become: how many administrative boundaries a typical service resolution crosses, what proportion of Internet communication is already name-based end-to-end, and how concentrated and fragile the infrastructure supporting that name-based
communication really is. 
This empirical foundation does not replace the architectural question of what should be done --- but it clarifies the stakes with quantitative evidence and underscores the
urgency of the problem that documented incidents alone cannot provide.

\section{Conclusion}
\label{sec:conclusion}

The Internet has undergone a fundamental operational transformation: from
an architecture in which IP addresses were the stable identifiers for communicating entities, to one in which DNS names serve that purpose, while the role of IP addresses has been reduced to ephemeral routing locators. 
This transformation was not intentionally designed; rather, it was the cumulative outcome of independent operational decisions, each reasonable on its own, that collectively replaced the original architecture's assumptions, without a coherent overall alternative.

The consequences are not theoretical. The indirection stack that bridges
name-based operation and address-based forwarding has become load-bearing
infrastructure, and its failure modes are already visible at scale. 
The outages documented in Section~\ref{sec:consequences} share a common
pattern: the address-based packet-delivery substrate was functional; the name-based
operational layer --- DNS steering and CDN orchestration --- was the point of failure as operations quietly converged on names.
The security infrastructure, fragmented across incompatible identifier spaces as documented in Section~\ref{sec:security}, compounds this fragility; its full architectural implications remain future work.

What the Internet community has not yet done is measure the full extent of this condition. How many administrative boundaries does a typical service resolution cross? What fraction of Internet traffic already flows exclusively over name-identified, TLS-secured channels? How concentrated are the backend dependencies that underlie the apparent diversity of CDN infrastructure? These questions are answerable---the measurement methods and vantage points exist---but the answers are not yet in hand. Until they are, the argument for architectural changes relies on incidents rather than systematic evidence.

Existing evidence suggests that the Internet has already become name-based in practice. The question is whether its architecture will catch up deliberately, or whether the gap will continue to widen---accumulating complexity, fragility, and risk---until the next wave of failures forces a change.

\bibliographystyle{ACM-Reference-Format}
\bibliography{refs}

\end{document}